%% file: WmassHCP2008.tex
\documentclass[slac_one]{revtex4}
\usepackage{graphicx}
\usepackage{subfigure}
\usepackage{fancyhdr}
\newcommand{\pythia}  {{\sc Pythia}}
\newcommand{\resbos}  {{\sc RESBOS}}

\newcommand{\wgrad}   {{\sc WGRAD}}
\newcommand{\cteqsixm}{{\sc CTEQ6M}}
\newcommand{\geant}   {{\sc GEANT}}

\newcommand{\gevcd} {\mbox{${\rm GeV}/c^2$}}
\newcommand{\mevcd} {\mbox{${\rm MeV}/c^2$}}

\newcommand{\cm}    {\mbox{${\rm cm}$}}

\newcommand{\ipb}   {\mbox{${\rm pb^{-1}}$}}
\newcommand{\ifb}   {\mbox{${\rm fb^{-1}}$}}

\newcommand{\Ups}      {\mbox{$\Upsilon(1S)$}}
\newcommand{\Jpsi}     {\mbox{$J/\psi$}}

\newcommand{\Zll}      {\mbox{$Z       \to \ell^+ \ell^-$}}

\newcommand{\Zee}      {\mbox{$Z       \to  e^+   e^-$}}
\newcommand{\Zmumu}    {\mbox{$Z       \to \mu^+  \mu^-$}}
\newcommand{\Upsmumu}  {\mbox{$\Ups    \to \mu^+  \mu^-$}}
\newcommand{\Jpsimumu} {\mbox{$J/\psi  \to \mu^+  \mu^-$}}

\newcommand{\Wenu}     {\mbox{$W \to  e    \nu$}}
\newcommand{\Wmunu}    {\mbox{$W \to  \mu  \nu$}}
\newcommand{\Wtaunu}   {\mbox{$W \to  \tau \nu$}}
\newcommand{\Wlnu}     {\mbox{$W \to  \ell \nu$}}

\newcommand{\sumet}  {\mbox{$\Sigma{E_T}$}}
\newcommand{\upara}  {\mbox{$u_{||}$}}

\newcommand{\mtr}  {\mbox{$m_T$}}
\newcommand{\mw}   {\mbox{$m_W$}}
\newcommand{\mh}   {\mbox{$m_H$}}
\newcommand{\mt}   {\mbox{$m_t$}}
\newcommand{\mz}   {\mbox{$m_Z$}}

\newcommand{\pt}   {\mbox{$p_T$}}
\newcommand{\ptl}  {\mbox{$p_T^l$}}
\newcommand{\ptnu} {\mbox{$p_T^\nu$}}

\newcommand{\metEq}  {{E}\!\!\!/_T }
\newcommand{\met}    {\mbox{$\metEq$}}

\newcommand{\ep}      {\mbox{$E/pc$}}

\newcommand{\lumiunits} {\mbox{$10^{30}~\rm{s}^{-1}~\cm^{-2}$}}

\newcommand{\dmzstat}  {\mbox{$\Delta m_Z^{\rm stat.}$}}
\newcommand{\dmwstat}  {\mbox{$\Delta m_W^{\rm stat.}$}}
\newcommand{\dmwscstat}{\mbox{$\Delta m_W^{\rm scale (stat.)}$}}

\newcommand{\ppbar}    {\mbox{$p\bar{p}$}}

\newcommand{\dzero}    {D\O}

\begin{document}

\title{{\small{Hadron Collider Physics Symposium (HCP2008),
Galena, Illinois, USA}}\\ 
\vspace{12pt}
W boson mass measurement at the Tevatron} 

%

\author{I. Bizjak (for the CDF Collaboration)}
\affiliation{
  University College London, Gower Street, LONDON, WC1E 6BT, UK
  \footnote{On leave from J. Stefan Institute, Jamova 39, Ljubljana, Slovenia.}
}
\begin{abstract}
  
  The mass of the $W$ boson is one of the least precisely measured
  parameters of the electroweak interaction. Confronted with other
  measurements of standard model parameters it can test the internal
  consistency of the Standard Model and can constrain the possible
  mass of the standard model Higgs boson. The CDF collaboration has
  published the current single most precise measurement of the $W$
  boson mass using $200~\ipb$ of CDF Run II \ppbar\ data, and an
  improved measurement with $2.4~\ifb$ of CDF Run II data is underway.
  Both measurements are described in detail in these proceedings.

\end{abstract}

\maketitle

\thispagestyle{fancy}


\input{text.tex}


%

\end{document}

%% file: text.tex
\section{INTRODUCTION} 

The standard model of particle physics (SM) has been very successful
in describing electroweak interactions. The mass of the $W$ boson is
one of the parameters of the electroweak theory, which at tree level
is fully determined by the relations to the other electroweak
parameters. The loop corrections to the $W$ boson mass calculation are
dominated by the loop containing the $b$ and $t$ quarks and could
receive a contribution from a loop with the yet-unconfirmed Higgs
boson, while further corrections can come from loops with possible
supersymmetric particles. Precision measurements of the $W$ boson mass
allow us to constrain the size of the possible loop corrections,
giving us valuable information on the range of possible Higgs masses.
The proceedings present one of the most precise single measurements of
the $W$ boson mass.

\section{THE $W$ BOSON MASS MEASUREMENT AT CDF}

The Tevatron experiments CDF and \dzero\ have long been involved in the
measurement of the $W$ boson mass, first pushing the precision below
$100~\mevcd$, and later published the combined average of the Run I
$W$ boson mass with an uncertainty of $59~\mevcd$~\cite{wmassrun1}.
In 2001 the upgraded CDF and \dzero\ detectors began taking data and the
size of the Run I dataset has long been surpassed. The CDF detector
uses its tracking and calorimetry capabilities to measure the $W$
boson mass both in \Wenu\ and \Wmunu\ decays.

\subsection{Measurement Strategy}

The $W$ bosons in \ppbar\ collisions are mainly produced in quark
anti-quark annihilation. In the environment of hadronic machines, only
\Wenu\ and \Wmunu\ decays can be reconstructed with the purity needed
for the $W$ boson mass measurement. The neutrino is not detected in
the CDF detector and its momentum cannot be fully kinematically
constrained, since the momentum of the incoming partons producing the
$W$ boson is not known, and the momentum along the beam-line cannot be
accurately measured. The momentum conservation of the measured
quantities can instead be applied in the transverse plane, since the
partons have negligible transverse momenta. The transverse momentum of
the neutrino (\ptnu) is thus inferred from \met, the missing
transverse momentum required to achieve the momentum balance in the
transverse plane. The so-called transverse mass is constructed out of
the kinematic quantities in the transverse plane:
%
\begin{equation}\label{eq:mt}
\mtr=\sqrt{2 \ptl \cdot \ptnu \cdot \left[1- \cos(\phi_\ell - \phi_\nu)\right]}~~~.
\end{equation}
%
The dependence of \mtr\ on \mw\ is determined using Monte Carlo (MC)
templates, which are created for different input \mw. The template
best corresponding to the data distributions is extracted by a binned
maximum-likelihood fit. The $W$ boson mass is also obtained from the
templates constructed for lepton transverse momentum \ptl\ and the
transverse momentum of the neutrino, \met. The final \mw\ value is a
weighted average of the three determinations. While the three
variables are highly correlated, they have different sensitivities to
boosts and systematic effects, hence their combination improves the
\mw\ determination.

For the desired accuracy of the \mw\ measurement the \mtr, \ptl, and
\met\ templates need to be generated using a far greater number of
events than would be feasible with full detector simulation. A fast
parameterized detector simulation is developed for this purpose and is
described in detail in Ref.~\cite{wmass}. The fast event generation
and detector simulation enables reliable estimates of systematic
uncertainties since we are able to produce several simulation samples
where the relevant parameters are varied within their uncertainty.

The main quantities in the event measured by the detector are the
momentum of the charged lepton and the hadronic recoil, the total
transverse momentum of the remaining particles, recoiling (in the
transverse plane) from the $W$ boson. The measurement of the recoil is
necessary to reconstruct the missing transverse momentum \met.
Recoiling particles come from gluon radiation of the incoming quarks,
and the underlying event processes. The fast event generation and
detector simulation have to be calibrated to predict these kinematic
variables to a part in $10^4$.

\subsection{Momentum and Energy Measurement Calibrations}
%
\begin{figure*}[t]
\centering
  \subfigure[]{
    \includegraphics[width=80mm]{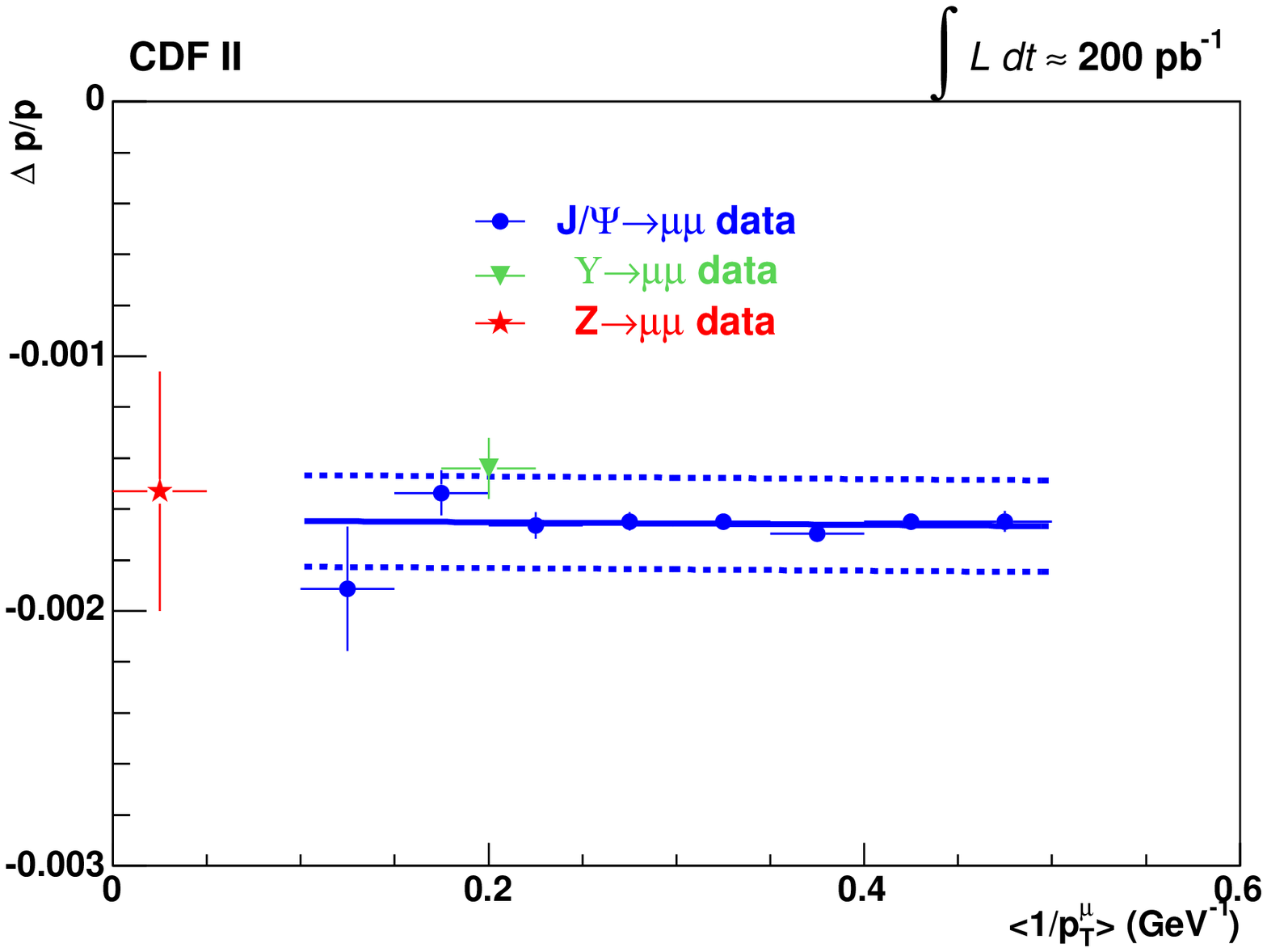}
    \label{dpoverp}
  }
  \subfigure[]{
    \includegraphics[width=80mm]{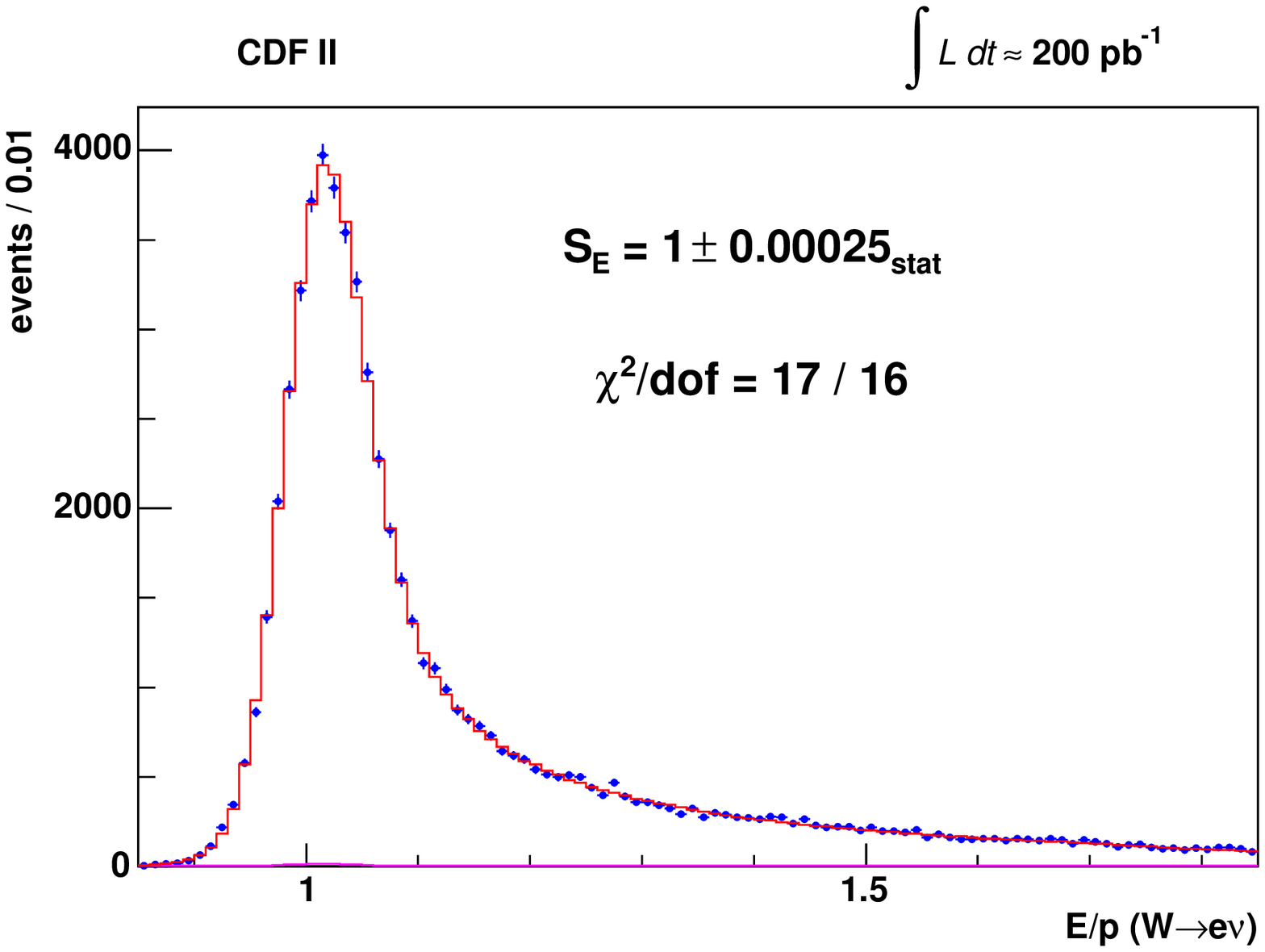}
    \label{eopdist}
  }
  \caption{
  (a)  
  The fractional momentum correction $\Delta p/p$ as a function of the
  mean inverse momentum of the muons from \Jpsimumu, \Upsmumu\ and
  \Zmumu\ decays. The full line is the best linear fit to the points, and the
  dotted lines mark the total (statistical and systematic) uncertainty
  of the fit.
  (b)
  The \ep\ distribution used for the calibration of the energy
  measurement.
}
\end{figure*}
%
The muon momentum is determined using the Central Outer Tracker (COT),
a gas-filled cylindrical drift chamber placed in a $1.4$~T magnetic
field. The internal alignment of the COT is performed using high-\pt\ 
cosmic rays. The fast simulation of the momentum measurement takes
into account the energy loss and scattering of particles in the
material, COT hit resolutions and the track-fitting algorithm. The
absolute momentum measurement scale and hit resolutions as well as the
number of simulated radiation lengths of traversed material are tuned
on data.

The energy of the electron is obtained from the measurement in the
electromagnetic calorimeter (EM), while the angular information of the
electron track is obtained with the COT. The simulation of the energy
measurement includes the nonlinearity of the calorimeter response, the
energy dependence of the energy measurement and the shower leakage out
of the EM calorimeter.

The calibration of the measured momenta and energies is one of the
most important aspects of the $W$ boson mass measurement, since
scaling the measured momenta directly shifts the fitted kinematic
distributions.  The absolute momentum scale is set by requiring that
the measured values of the precisely known masses of the $Z$, $J/\psi$
and $\Ups$ resonances in the decays to a pair of muons correspond to
their PDG values~\cite{pdg2006}. Figure~\ref{dpoverp} shows that the
fractional momentum scale corrections obtained from different
invariant mass fits are consistent with each other. Combining the
calibration using all three particle masses enables momentum
determination with an accuracy of 0.021\%.

Taking advantage of the high \Jpsimumu\ and \Upsmumu\ statistics, the
measured momentum can be calibrated more accurately than the energy
measured in the calorimeter. The momentum calibration is hence
transferred to the energy measurement using the \ep\ distribution
(Figure~\ref{eopdist}). We adjust the absolute energy scale on the
peak of the \ep\ distribution: the tuning is done in several
transverse energy bins to account for non-linear calorimeter response.
The number of radiation lengths of material used in the energy loss
simulation is tuned to match the radiative tail of the \ep\ 
distribution.

The calibration yields a \Zee\ mass measurement of $M_Z = 91190 \pm
67_{\rm stat}~\mevcd$, in very good agreement with the world average
($91187.6 \pm 2.1~\mevcd$~\cite{pdg2006}). The calorimeter calibration is
further improved by combining the tunes of the \ep\ and the \Zee\ 
invariant mass distributions. Similar to the scale, the resolution of
the energy measurement is tuned on the widths of the \ep\ and the
\Zee\ mass distributions.

\subsection{Hadronic Recoil Calibration}

Simulation of the reconstructed transverse momentum of the neutrino,
i.e. the missing transverse momentum \met, requires an accurate
simulation of the hadronic recoil, the total transverse momentum of
all particles except that of the two prompt leptons from the $W$ boson
decay. It is measured from all the energy depositions in the
calorimeter, except those associated with the primary charged lepton.
These depositions are caused by particles from initial state QCD
radiation, the hadronization of the \ppbar\ remnants not partaking in
the boson creation, multiple collisions from the same bunch crossing
and from initial and final-state photons.
%
\begin{figure*}[t]
  \centering
  \subfigure[]{
    \includegraphics[width=80mm]{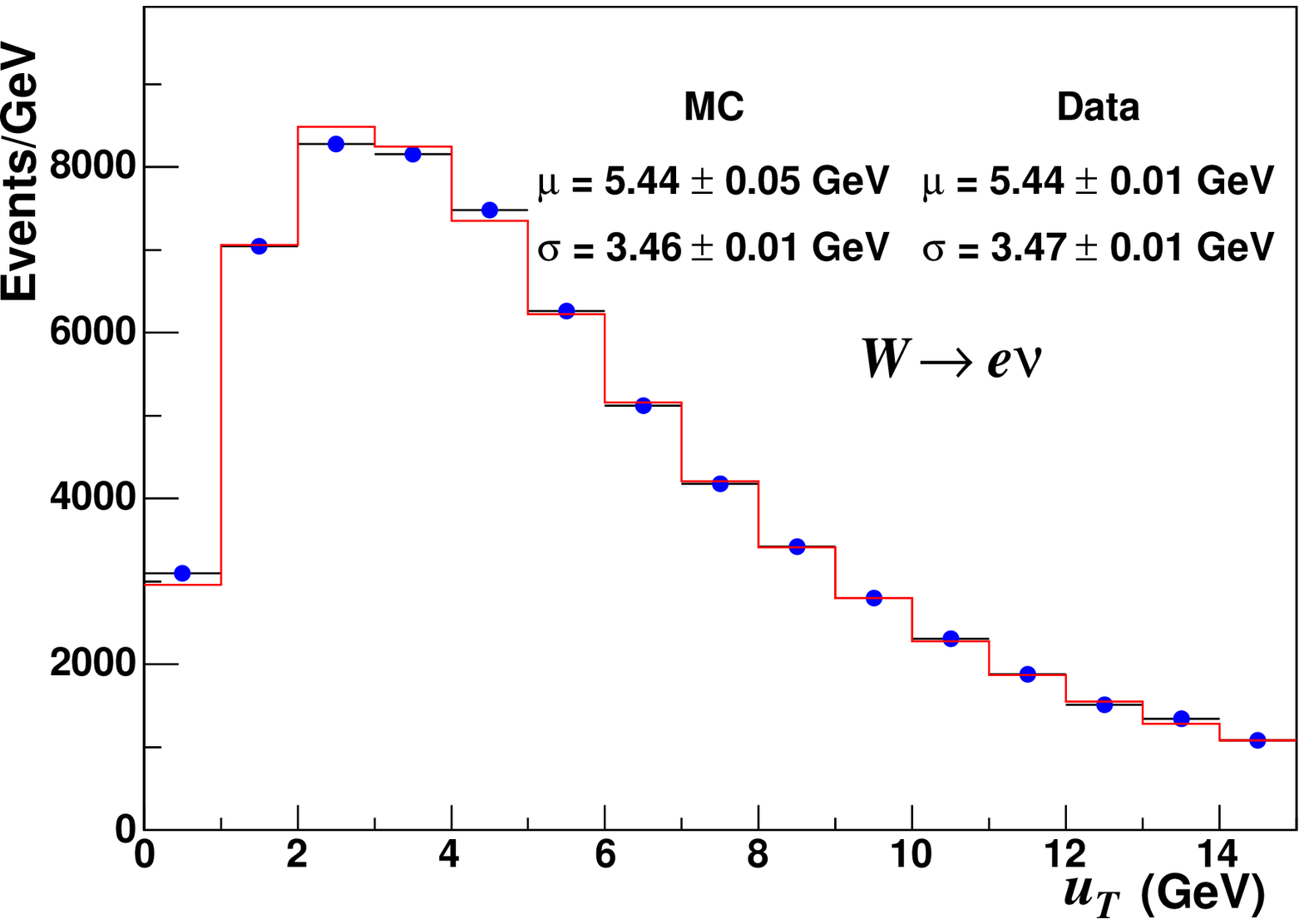}
    \label{weut200pb}
  }
  \subfigure[]{
    \includegraphics[width=80mm]{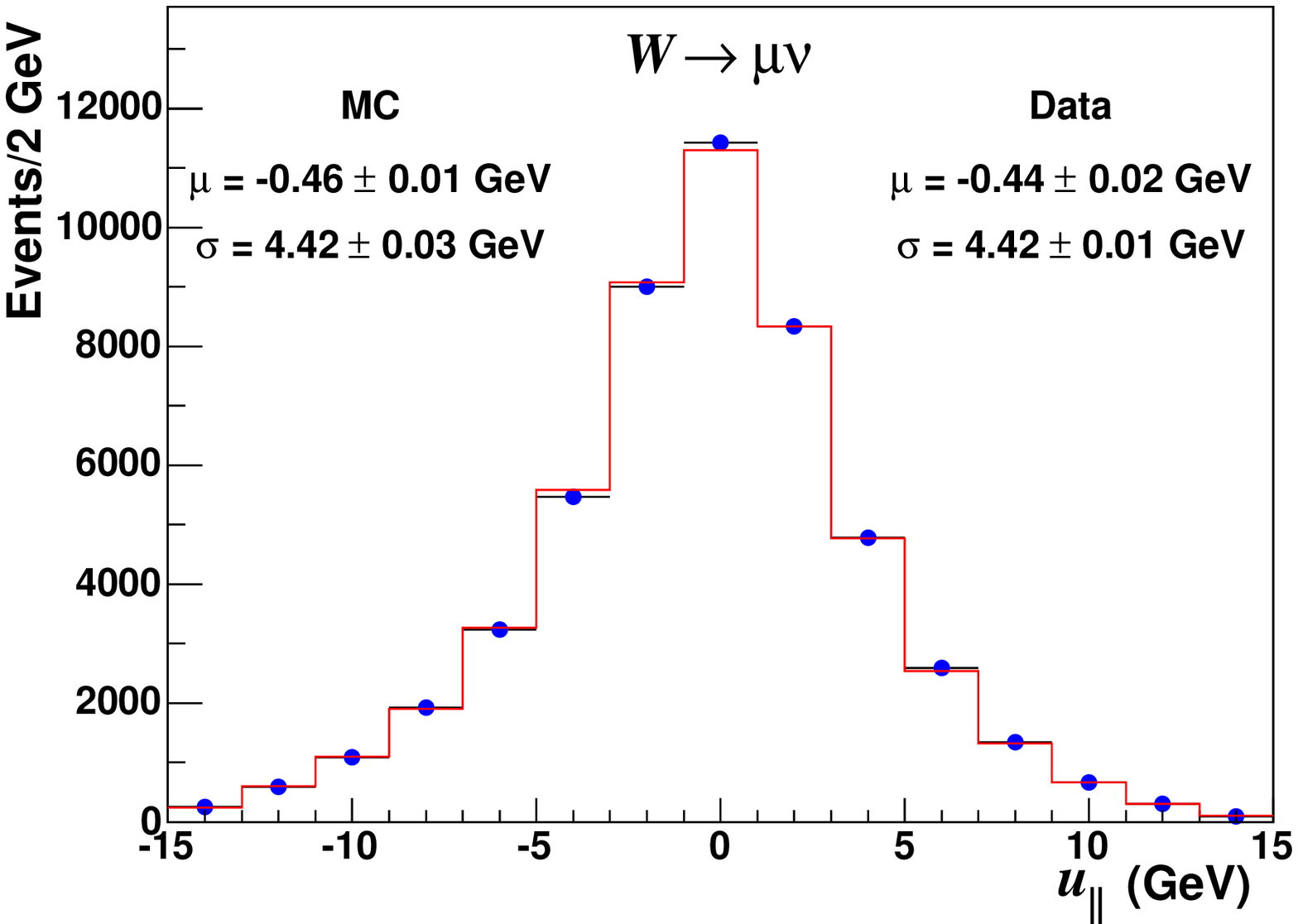}
    \label{wmupar200pb}
  }

  \caption{
    (a) The comparison of the simulated magnitude of the recoil with
    data, for \Wenu\ candidates. (b) The comparison of the
    simulated distribution of the recoil component in the direction of
    the charged lepton with data, for \Wmunu\ candidates.}
\end{figure*}
%
The contribution of different components to the magnitude and spread
of the recoil distributions is parameterized and the parameterization
tuned on the \Zll\ decays and minimum bias data (for the underlying
event). The comparison of the data and the simulated magnitude of the
recoil (Figure~\ref{weut200pb}) and \upara, the component of
the recoil in the lepton direction (Figure~\ref{wmupar200pb})
shows good agreement. The \mw\ measurement is specially sensitive to
\upara\ as it directly affects the calculation of \mtr.

\subsection{Generation of $W$ boson production and decay}

The dedicated parameterized Monte Carlo simulation has to model both
initial state gluon radiation, which is responsible for a non-zero $W$
boson \pt, and the QED radiation of the final state lepton(s).  The
$W$ boson \pt\ spectrum is modelled with \resbos~\cite{resbos} and its
parameterization of the non-perturbative low \pt\ region is tuned on the
dilepton \pt\ distribution in the \Zll\ decays. The QED
corrections of one-photon emission are simulated using
\wgrad~\cite{wgrad}. The photon energies are increased by 10\% (with an
absolute uncertainty of 5\%) to account for additional energy loss due
to two-photon radiation which is not included in the \wgrad\
calculation.

Systematic uncertainties due to the limited knowledge of the
non-perturbative QCD parameters determining the $W$ \pt\ spectrum and
the simulation of the radiation of more than one photon were estimated
(see Table~\ref{uncertainty}).

Parton distribution functions (PDFs) are used to calculate the
kinematics of the $W$ boson decay products. The \cteqsixm~\cite{cteq}
PDFs and their error sets are used for the generation of templates and
the PDF uncertainty estimation~(Table~\ref{uncertainty}).

\subsection{Backgrounds}

Due to imperfect detector measurement and coverage, several background
processes can mimic a \Wlnu\ decay, thus passing the $W$ event selection.
We estimate the amounts and shapes of the dominant background
contributions and add them to the simulation templates when performing
the \mw\ fits.

The main backgrounds in the \Wenu\ channel are \Zee\ and \Wtaunu\ 
events, and events where a hadronic jet is mis-reconstructed as a
lepton (`multi-jet background'). We model \Zee\ and \Wtaunu\ events
using \pythia~\cite{pythia} and the full \geant\ simulation of the CDF
detector. The multi-jet background normalizations are found by fitting
the low \met\ distribution, where this background dominates.

The main backgrounds in the \Wmunu\ channel are \Zmumu\ and \Wtaunu\
events, cosmic rays, decays of low momentum pions or kaons into muons
in the COT, and hadronic jets, penetrating into the muon chambers or
decaying to muons. The \Zmumu\ and \Wtaunu\ events are modelled using
\pythia\ and the full detector simulation. The contribution of low
momentum meson decays is found by fitting the high tail of the track
fit $\chi^2$ distribution where this background is large.

\section{$W$ BOSON MASS FITS AND THE RESULT}

\begin{figure*}[t]
  \centering
  \subfigure[]{
    \includegraphics[width=80mm]{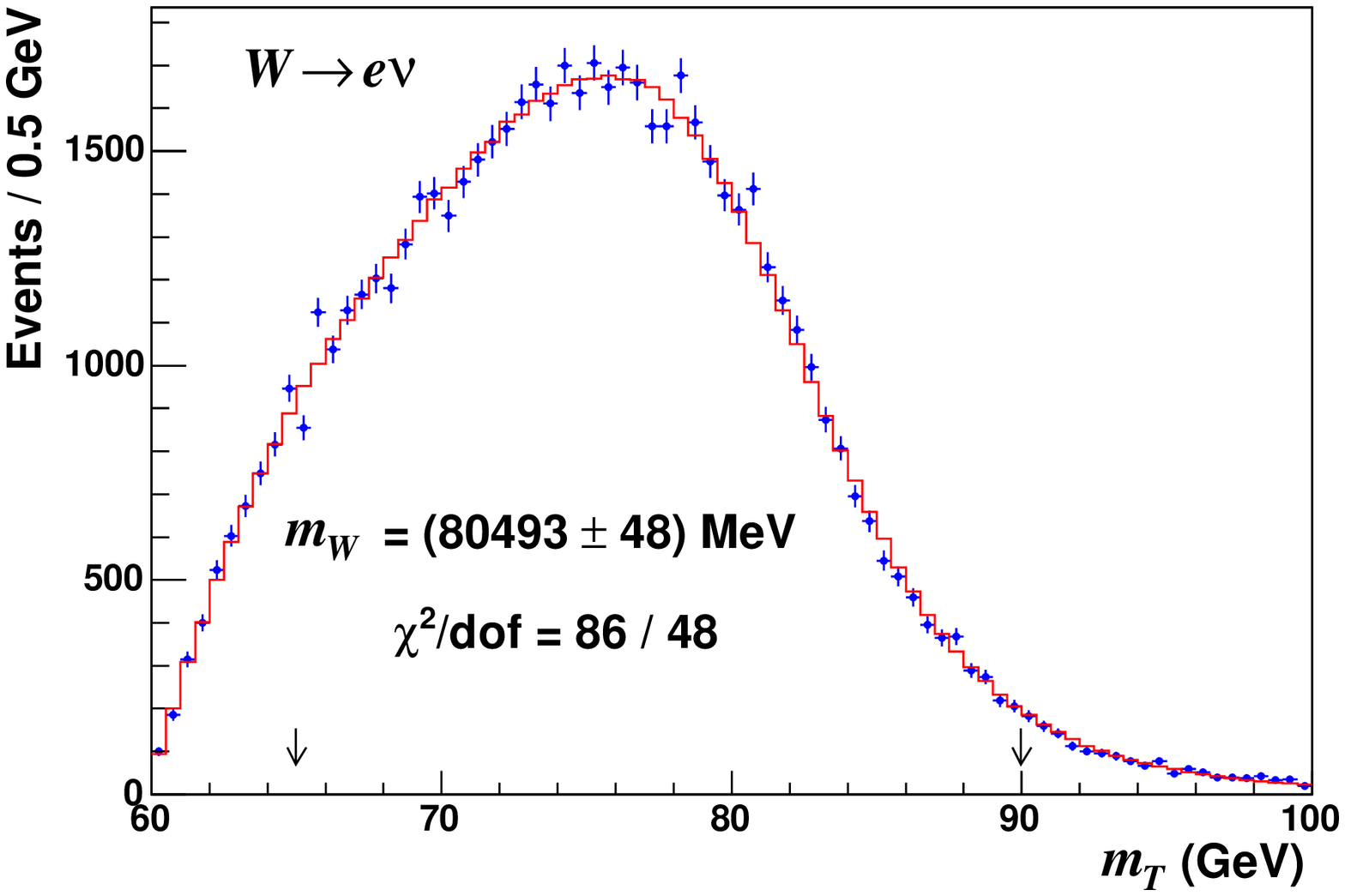}    
  }
  \subfigure[]{
    \includegraphics[width=80mm]{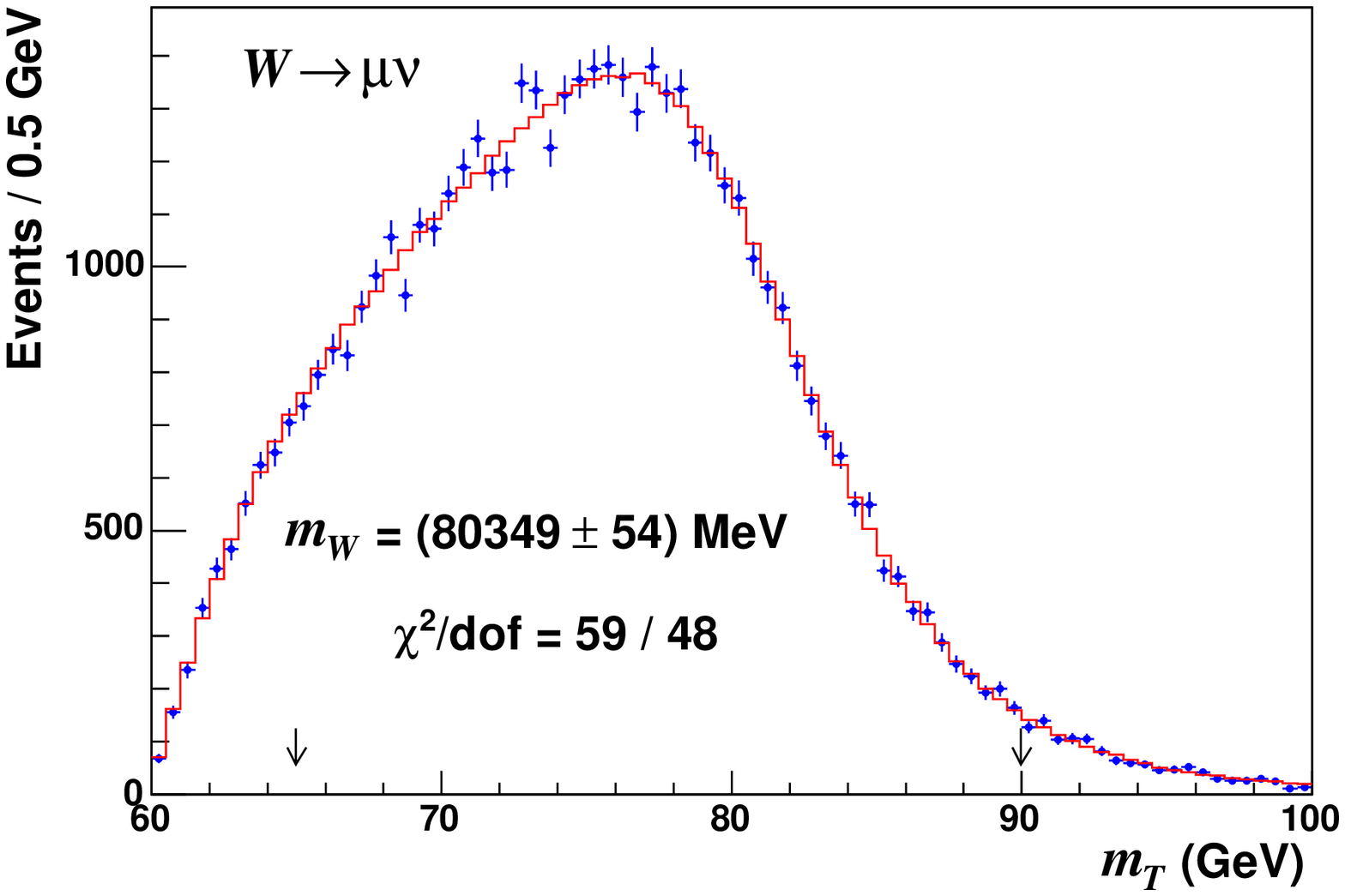}    
  }
\caption{
  Transverse mass fit in the \Wenu\ (a) and \Wmunu\ decay channel
  (b). The uncertainty given is statistical only. The arrows
  indicate the limits of the fit range.}
\label{results}
\end{figure*}

Finally, the $W$ boson mass is determined separately for the electron
and muon channels, in the fits to the three kinematic distributions
(\mtr, \ptl\ and \met). The results of the fits are summarized in
Table~\ref{fits}. The two transverse mass fits are shown in
Figure~\ref{results}.
\begin{table}[hb] 
\vspace{0.4cm}
\begin{center}
\begin{tabular}{|c|cc|}
\hline
{} &{} &{}\\[-1.5ex]
Distribution & \mw\ (\mevcd) & ~~$\chi^2$/dof~~ \\[1ex]
\hline
{} &{} &{} \\[-1.5ex]
$m_T(\Wenu)$       & ~$80493\pm 48_{\rm stat}\pm 39_{\rm syst}$ & 86/48 \\[1ex]       
$p_T^e (\Wenu)$    & ~$80451\pm 58_{\rm stat}\pm 45_{\rm syst}$ & 63/62 \\[1ex]     
$p_T^{\nu}(\Wenu)$ & ~$80473\pm 57_{\rm stat}\pm 54_{\rm syst}$ & 63/62 \\[1ex] 
\hline
{} &{} &{} \\[-1.5ex]
$m_T(\Wmunu)$      & ~$80349\pm 54_{\rm stat}\pm 27_{\rm syst}$ & 59/48 \\[1ex]
$p_T^\mu (\Wmunu)$ & ~$80321\pm 66_{\rm stat}\pm 40_{\rm syst}$ & 72/62 \\[1ex]
$p_T^{\nu}(\Wmunu)$& ~$80396\pm 66_{\rm stat}\pm 46_{\rm syst}$ & 44/62 \\[1ex]
\hline
\end{tabular}
\end{center}
\caption{
  Fit results with total systematic and statistical uncertainties from
  all three fit distributions used to extract \mw.}
\label{fits}
\end{table}
%
%
The breakdown of the systematic uncertainties for the example of the
\mtr\ fit is given in Table~\ref{uncertainty}.
%
\begin{table}[hb]
\vspace{0.4cm}
\begin{center}
\begin{tabular}{|l|c|c|c|}
\hline
{} &{} &{} &{} \\[-1.5ex]
Syst. uncertainty (\mevcd)~~ & ~~\Wenu~~ & ~~\Wmunu~~ & ~~Common~~ \\[1ex]
\hline
{} &{} &{} &{}\\[-1.5ex]
Lepton: Scale       & 30 & 17 & 17\\[1ex]
Lepton: Resolution  &  9 &  3 &  0\\[1ex]
Recoil: Scale       &  9 &  9 &  9\\[1ex]
Recoil: Resolution  &  7 &  7 &  7\\[1ex]
Selection           &  3 &  1 &  0\\[1ex]
Lepton Removal      &  8 &  5 &  5\\[1ex]
Backgrounds         &  8 &  9 &  0\\[1ex]
$p_T(W)$ Model      &  3 &  3 &  3\\[1ex]
PDFs                & 11 & 11 & 11\\[1ex]
QED radiation       & 11 & 12 & 11\\[1ex]
\hline
{} &{} &{} &{}\\[-1.5ex]
Total Systematics   & 39 & 27 & 26\\[1ex]
\hline
{} &{} &{} &{}\\[-1.5ex]
Total Uncertainty   & 62 & 60 & 26\\[1ex]
\hline
\end{tabular}
\end{center}
\caption{
  Systematic and total \mw\ uncertainties for the transverse mass fits 
  in \Wenu\ and \Wmunu\ channels. The correlated part of each systematic 
  uncertainty between the two channels is given in the last column. 
  All values in \mevcd.
}
\label{uncertainty}
\end{table}
%

The six $W$ boson mass fits are combined, taking into account the
correlations, to give $\mw=80413 \pm 34_{\rm stat} \pm 34_{\rm
  syst}~\mevcd$.  With a total uncertainty of $48~\mevcd$, this
measurement is the most precise single measurement of \mw\ to date.
This result increases the world average $W$ boson mass to $\mw = 80398
\pm 25~\mevcd$~\cite{lepwmass}, reducing its total uncertainty by
$15\%$.
%
\begin{figure*}[t]
  \centering
  \includegraphics[width=80mm]{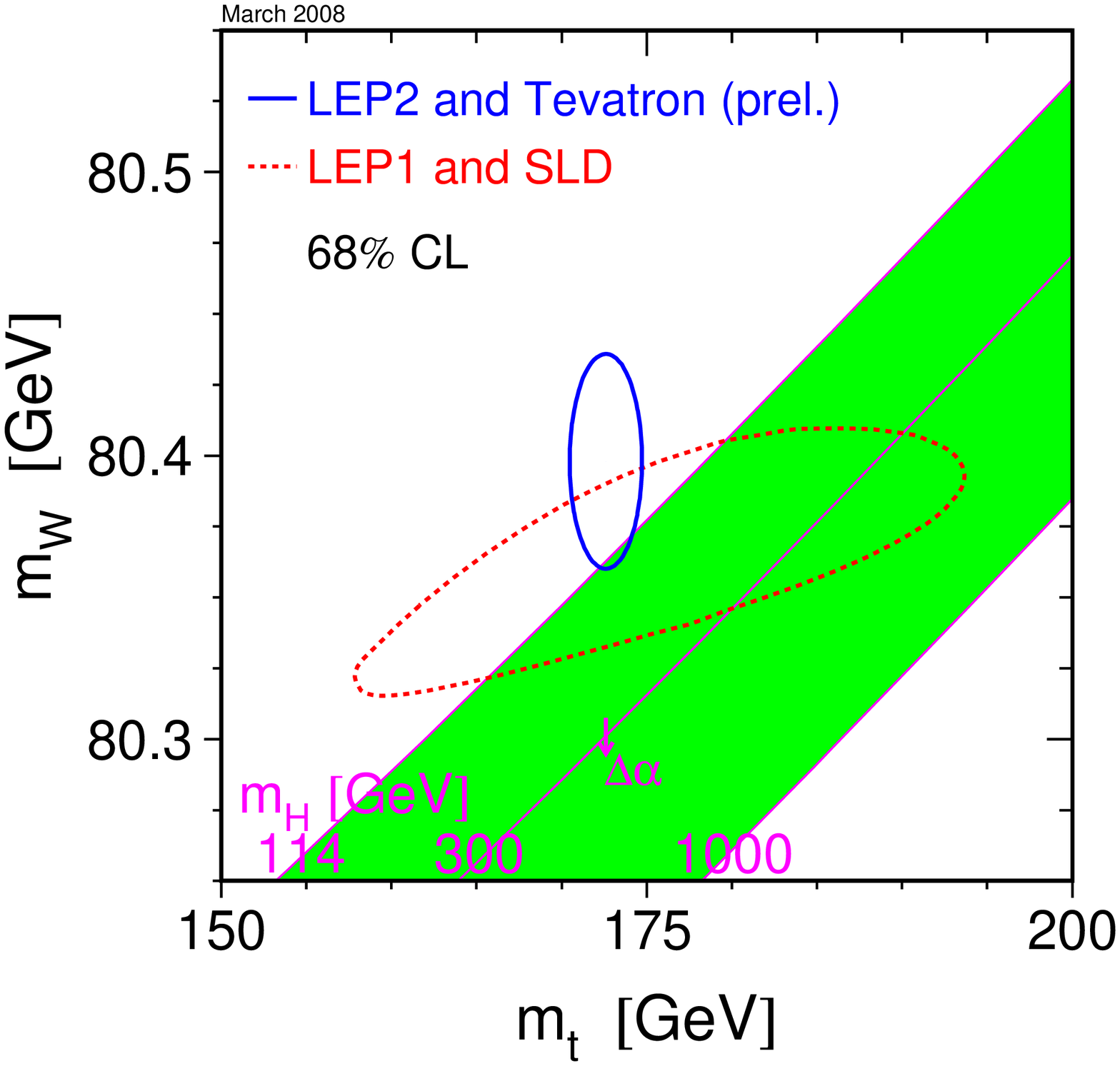}
  \caption{ 
    The constraint on the Higgs boson mass from direct \mw\ and \mt\ 
    measurements at 68\% confidence level (CL). The band shows the SM
    relationship of the two masses as a function of \mh\, which can be
    read from the abscissa.}\label{higgs}
\end{figure*}
%

A global electroweak fit with the new \mw\ average together with the
current best measurement of the top quark mass, $\mt = 172.6 \pm
1.4~\gevcd$~\cite{topmass}, constrains the possible Higgs boson mass
to $\mh=86^{+36}_{-27}~\gevcd$~\cite{lepwmass}. The corresponding
$95\%$ confidence level upper limit on the Higgs boson mass is
$160~\gevcd$~\cite{lepwmass}. Figure~\ref{higgs} shows the comparison
of the \mw\ and \mt\ measurements with their constraint on the Higgs
boson mass. The \mw\ and \mt\ units of the plot reflect how the
uncertainty on both masses propagates to the uncertainty on the
predicted \mh. To achieve the same level of constraint from $W$ boson
and top-quark mass measurements requires their uncertainties to
satisfy $\Delta m_W \approx 0.006 \cdot \Delta m_t$. It can be seen
from Figure~\ref{higgs} that currently the uncertainty of the \mw\ 
measurement is the limiting factor, thus motivating further work.

The total uncertainty of the published CDF measurement is smaller than
it was projected for $200~\ipb$ from the Run I CDF and \dzero\ 
measurements, mainly due to the inclusion of the \Ups\ and \Jpsi\ 
resonances in the momentum calibration and the usage of the \ep\ 
distribution for the energy measurement calibration.  Further
improvements in the detector model and the production and decay model
(for example QED radiative corrections) are likely to reduce other
systematic uncertainties as well. The CDF collaboration therefore
predicts that the measurement of the $W$ boson mass with a precision
better than that of the current world average ($25~\mevcd$) is
possible with the CDF data already in hand.

\section{FIRST LOOK AT THE \boldmath{2.4~\ifb} CDF RUN-II DATA}

The CDF collaboration has now collected well over ten times more data
than were used in the published analysis~\cite{wmass}. Most of the
systematic uncertainties in the published measurement
(Table~\ref{uncertainty}) are limited by the statistics of the
calibration samples. The sensitivity of a precision measurement to
much higher instantaneous luminosities and the fact that the data from
a much longer period of data-taking have to be calibrated, prevent a
straight-forward inclusion of more data.
%
%

The CDF collaboration has selected a data sample corresponding to
approximately $2.4$~\ifb\ that is most suitable for the improved \mw\ 
measurement and have already performed several calibrations with the
data (due to different triggering requirements the \Wmunu\ and \Zmumu\ 
decay samples correspond to approximately $2.3$~\ifb, as stated on the
plots). The following preliminary figures show that the increased
luminosity and range of the data samples have neither significantly
deteriorated the quality of the collected data nor our ability to
measure the quantities of interest.
%
\begin{figure*}[t]
  \centering
  \subfigure[]{
    \includegraphics[width=80mm]{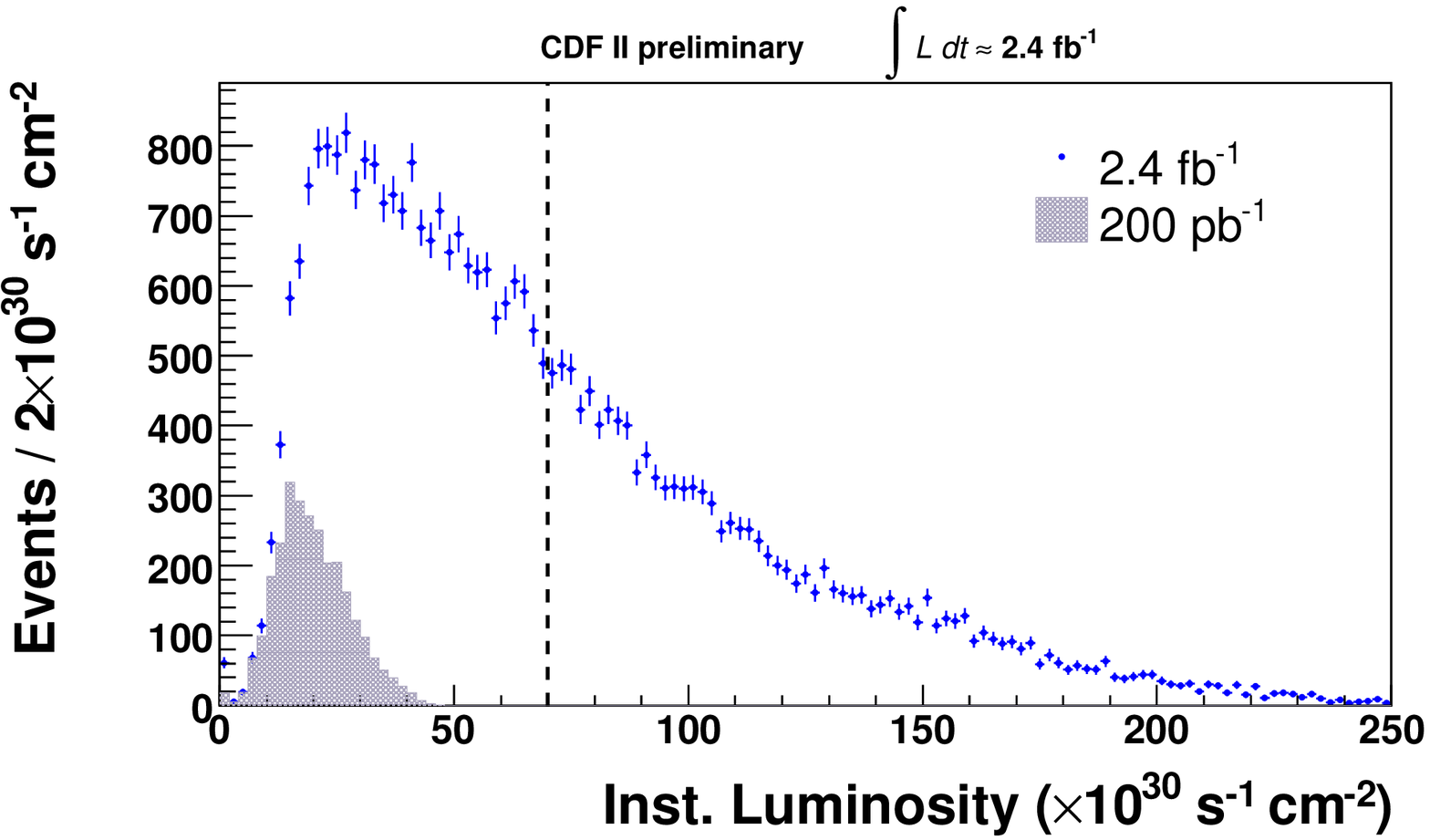}
    \label{lumi}
  }
  \subfigure[]{
    \includegraphics[width=80mm]{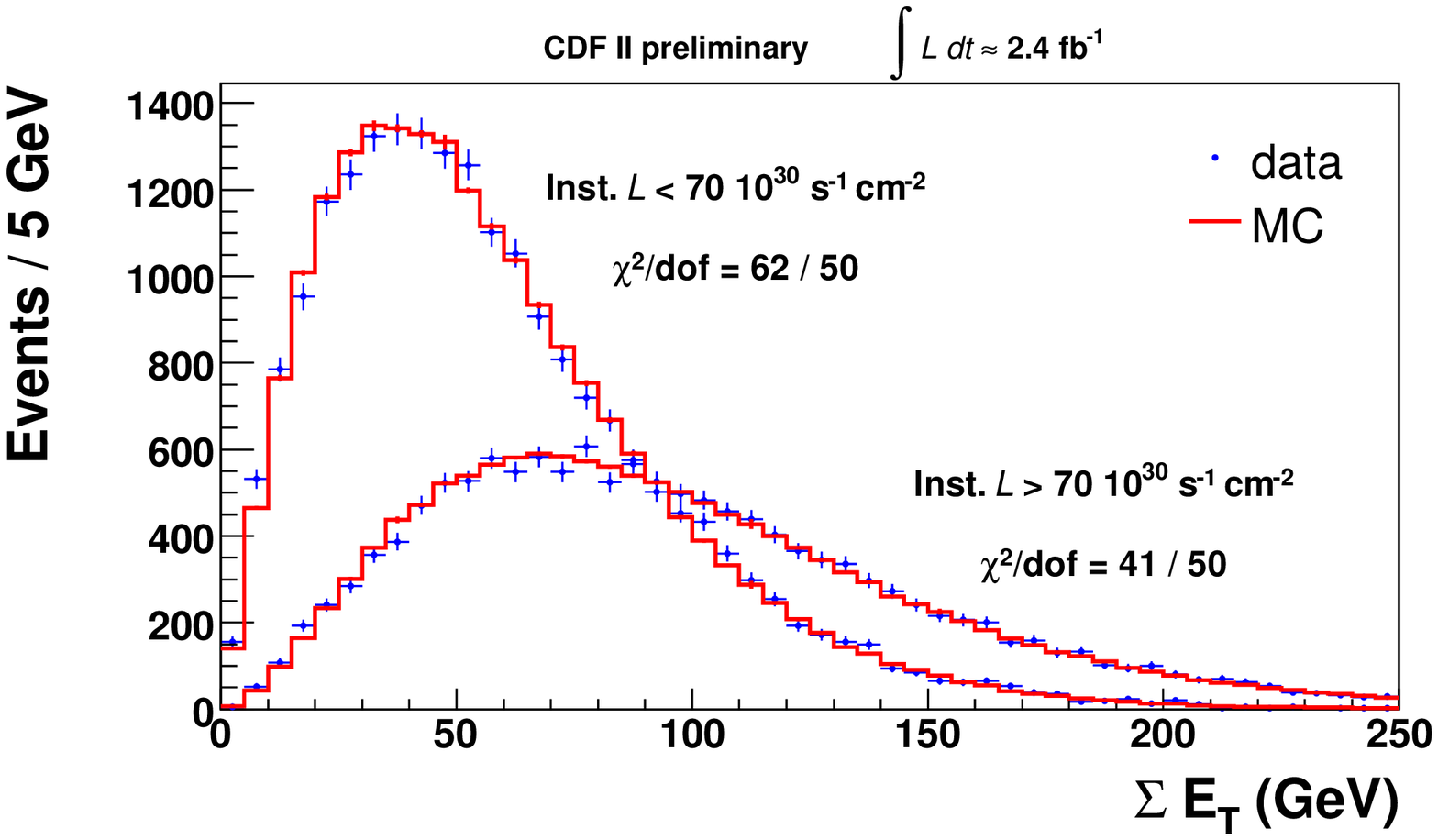}    
    \label{sumet}
  }

  \caption{
    (a) the comparison o the instantaneous luminosity distributions
    for the published analysis~\cite{wmass} and the preliminary
    $2.4$~\ifb\ data. (b) The distribution of \sumet\ for events at
    instantaneous luminosity greater and smaller than the average
    instantaneous luminosity of the preliminary $2.4$~\ifb\ dataset.
    The plots are for \Zee\ events.}
\end{figure*}
%
\subsection{Instantaneous Luminosity}

Figure~\ref{lumi} shows the comparison of the instantaneous luminosity
distributions for the published analysis and the preliminary
$2.4$~\ifb\ data.  Higher average instantaneous luminosities and a
much wider distribution of luminosity values affect the modelling of
multiple \ppbar\ interactions and backgrounds.  The recoil
distributions are the ones most severely altered and the recoil
simulation now includes the instantaneous luminosity in its
parameterization. The dependence of the hadronic recoil on the
instantaneous luminosity can be seen in the distributions of \sumet,
the total recoil energy deposit in the calorimeter.
Figure~\ref{sumet} shows the comparison of the data and the simulated
distribution of \sumet\ for events with instantaneous luminosity
smaller and greater than $70\times\lumiunits$ (this value roughly
corresponds to the average instantaneous luminosity of the whole
sample) for \Zee\ events. The agreement shows that the
parameterizations are able to capture the dependence of the total
recoil energy deposition on luminosity.

\subsection{Statistical scaling of uncertainties}

The degradation of the detector performance and the changes introduced
to the detector over time are systematic effects that can spoil the
scaling of measurement uncertainties with statistics. Special care has
been taken with the data calibrations to keep the quality of the new
data comparable to the one in the published analysis.

Figures~\ref{jpsiupsfit} and~\ref{zmmfit} show the invariant mass
distributions used for the calibration of the momentum measurement.
The fit for the momentum scale in \Jpsimumu\ decays is done in bins of
average curvature, the example plot shown, Figure~\ref{jpsifit} is for
the bin corresponding to the muons with the highest reconstructed
momenta.  Figures~\ref{zeefit} and~\ref{eop} show the kinematic
variables used for the calibration of the energy measurement. The
invariant mass of the candidate electrons from the \Zee\ decays, using
only the information from the COT, is shown in Figure~\ref{zeetrack}.
Due to the significant energy loss suffered by the electrons as they
pass through the detector material this distribution has worse $Z$
boson mass resolution than the invariant mass constructed from
calorimeter information. Good agreement between data and simulation
gives confidence in the simulation of the passage of electrons through
the detector.  Figure~\ref{wmtfit} shows the \mtr\ distributions for
data and simulation, used in the \mw\ fit.

\begin{figure*}[t]
  \centering
  \subfigure[]{
    \includegraphics[width=80mm]{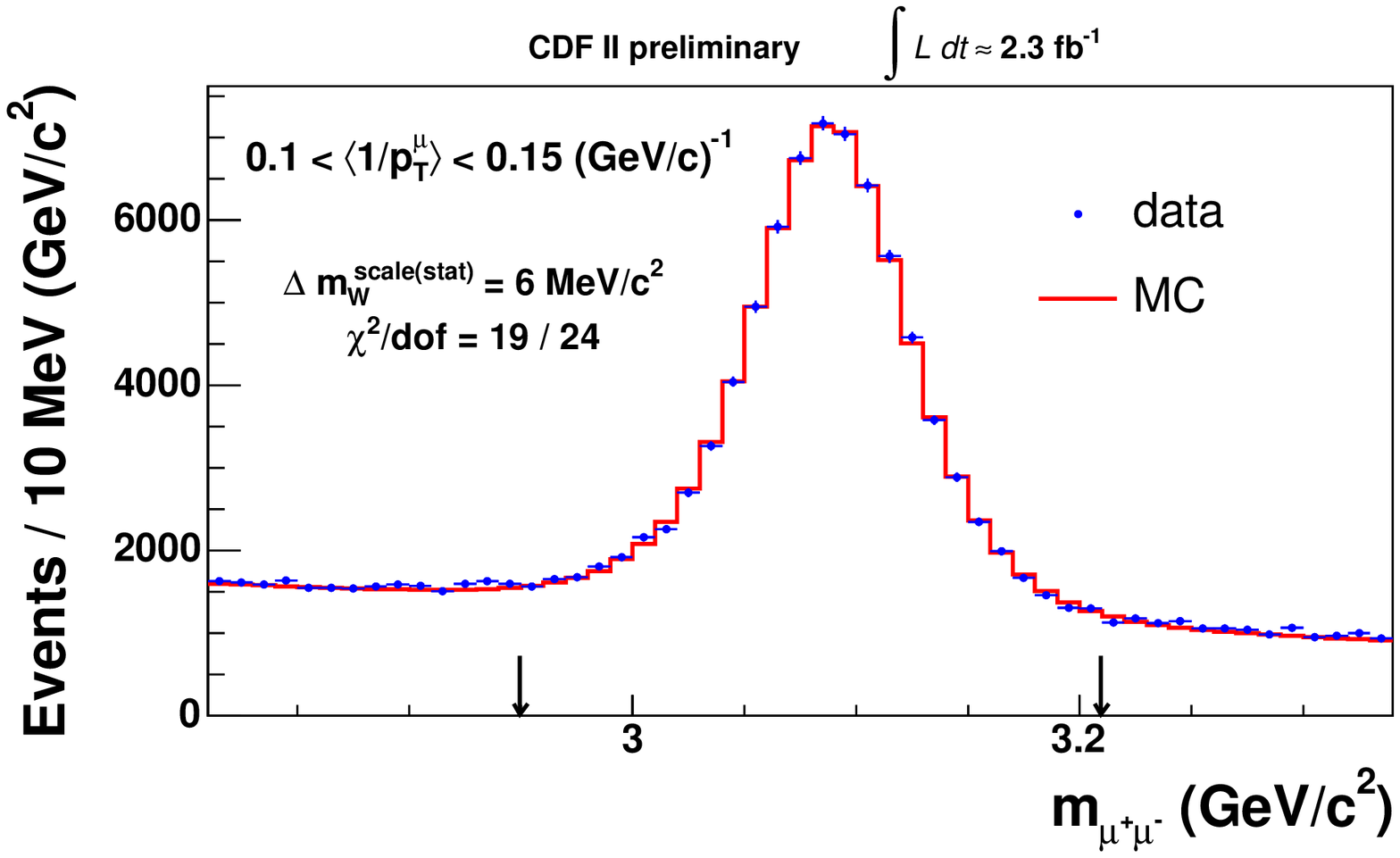}  
    \label{jpsifit}
  }
  \subfigure[]{
    \includegraphics[width=80mm]{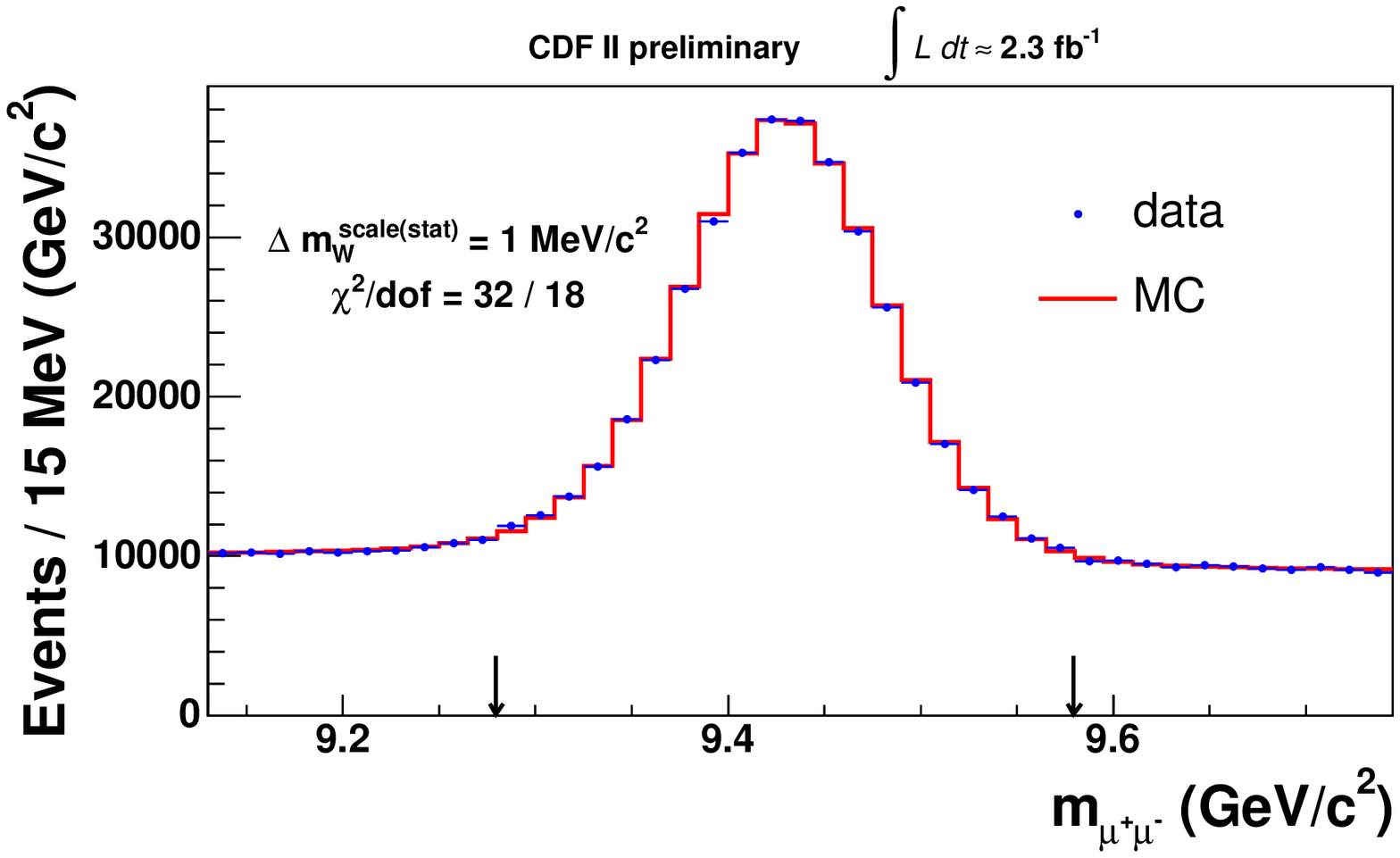}
    \label{upsfit}
  }
  \caption{
    The invariant dimuon mass for \Jpsimumu\ (a) and \Upsmumu\ 
    decays (b). The plot for \Jpsimumu\ is for muons with highest
    average momenta. The arrows indicate the limits of the fitting
    window.}
  \label{jpsiupsfit}
\end{figure*}

\begin{figure*}[t]
  \centering
  \subfigure[]{
    \includegraphics[width=80mm]{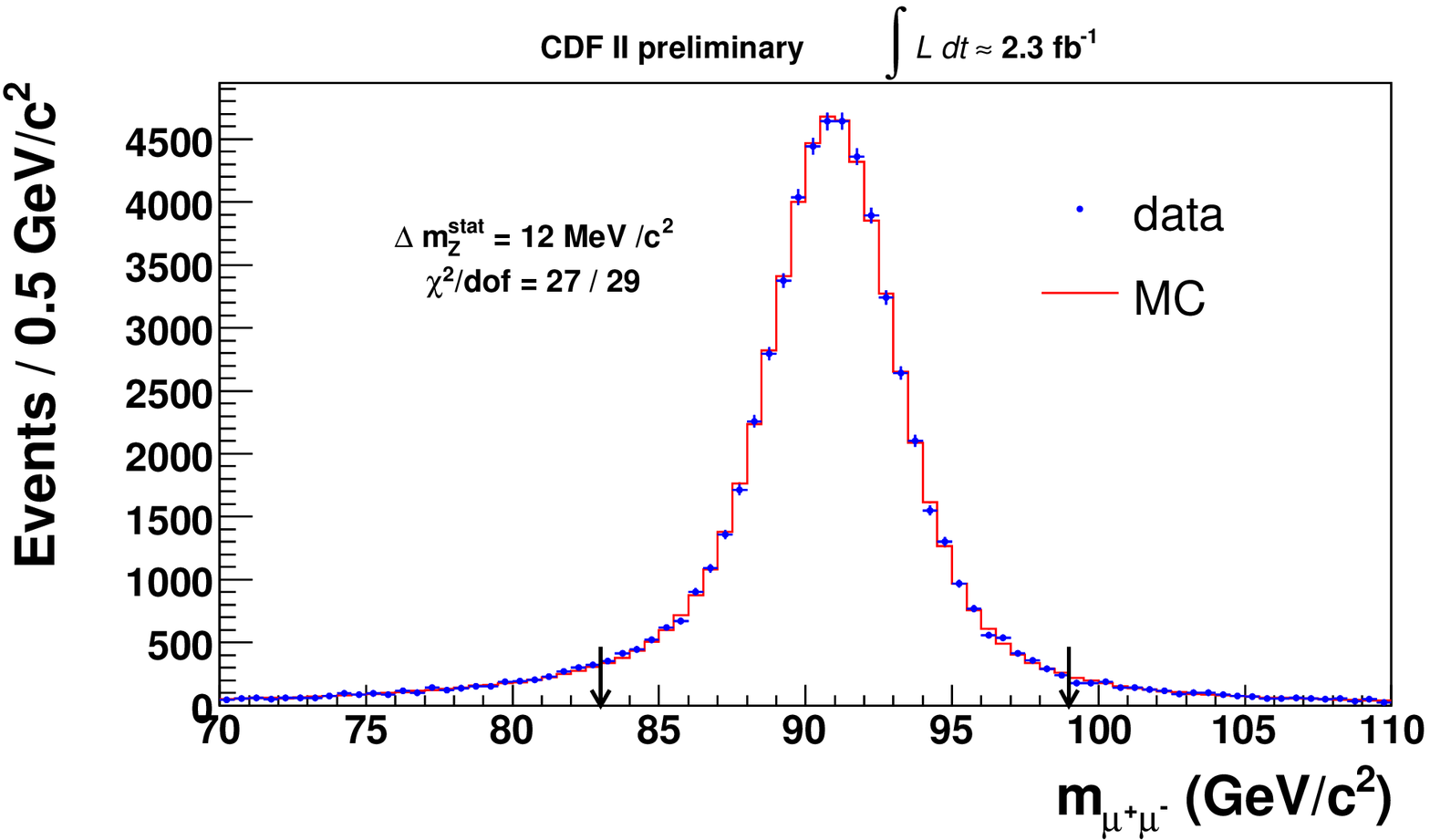}
    \label{zmmfit}
  }
  \subfigure[]{
    \includegraphics[width=80mm]{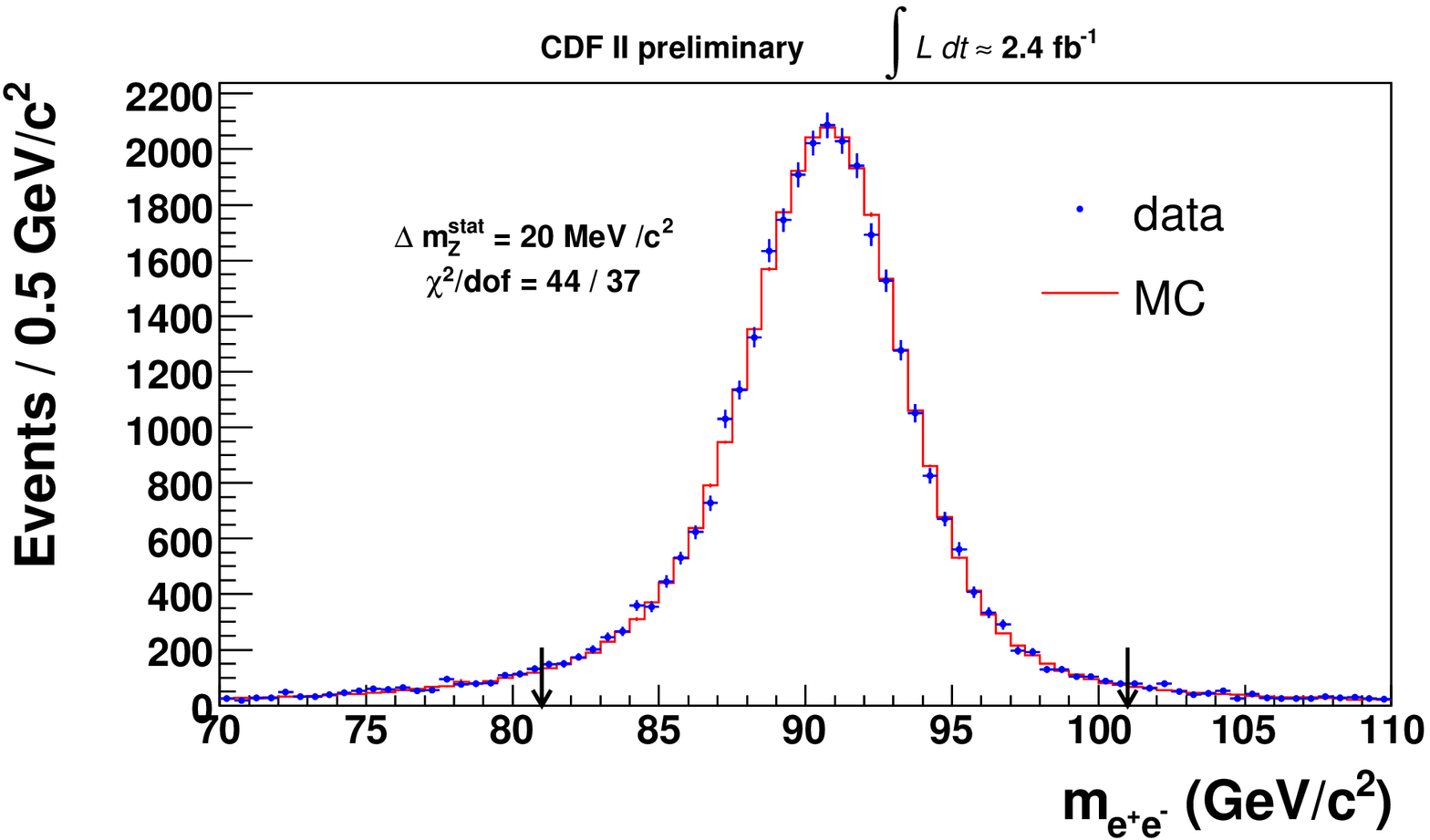}
    \label{zeefit}
  }
  \caption{The invariant dimuon mass for \Zmumu\ (a) and \Zee\ 
    decays (b). The arrows indicate the limits of the fitting
    window.}
\end{figure*}

\begin{figure*}[t]
  \centering
  \subfigure[]{
    \includegraphics[width=80mm]{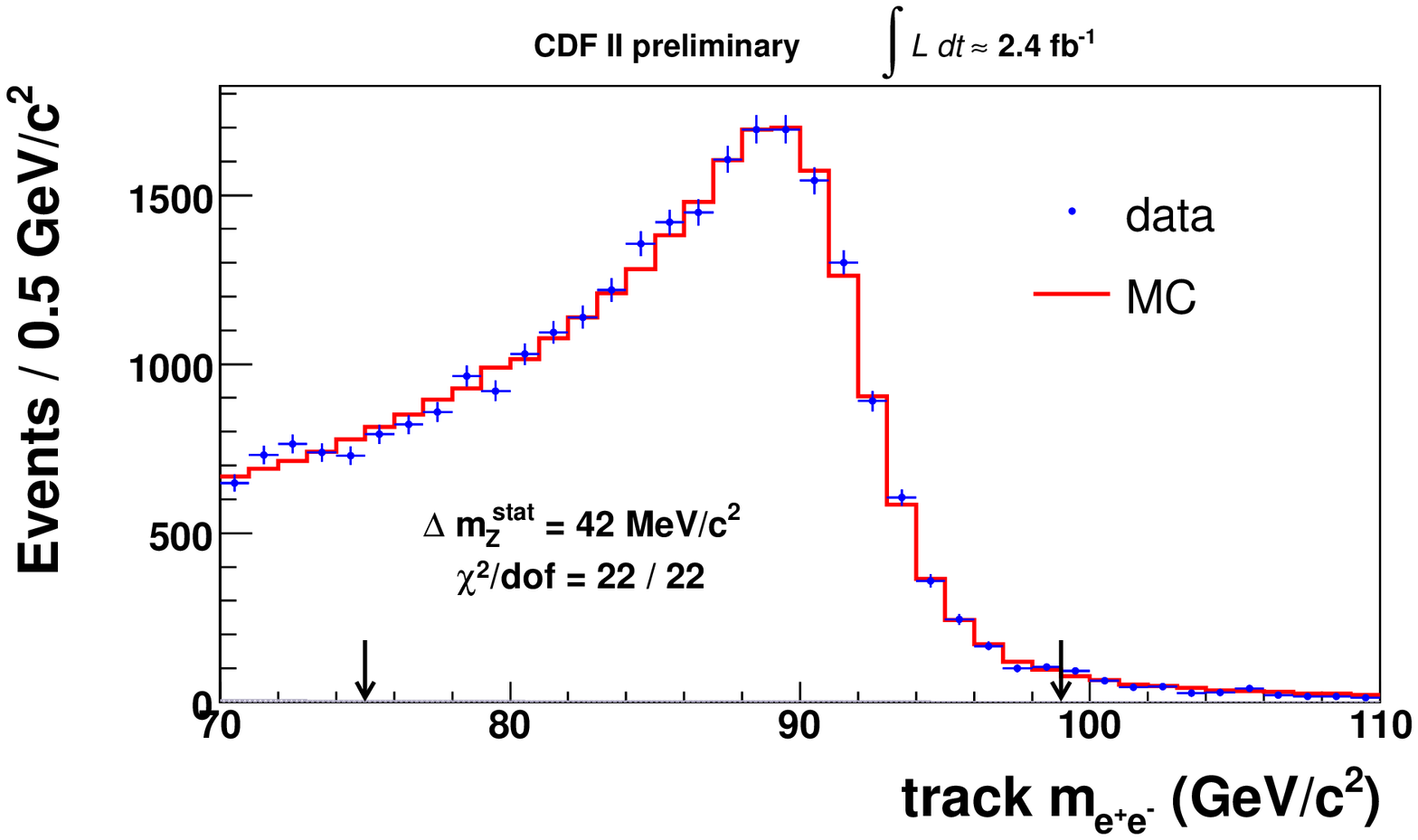}   
    \label{zeetrack}
  }
  \subfigure[]{
    \includegraphics[width=80mm]{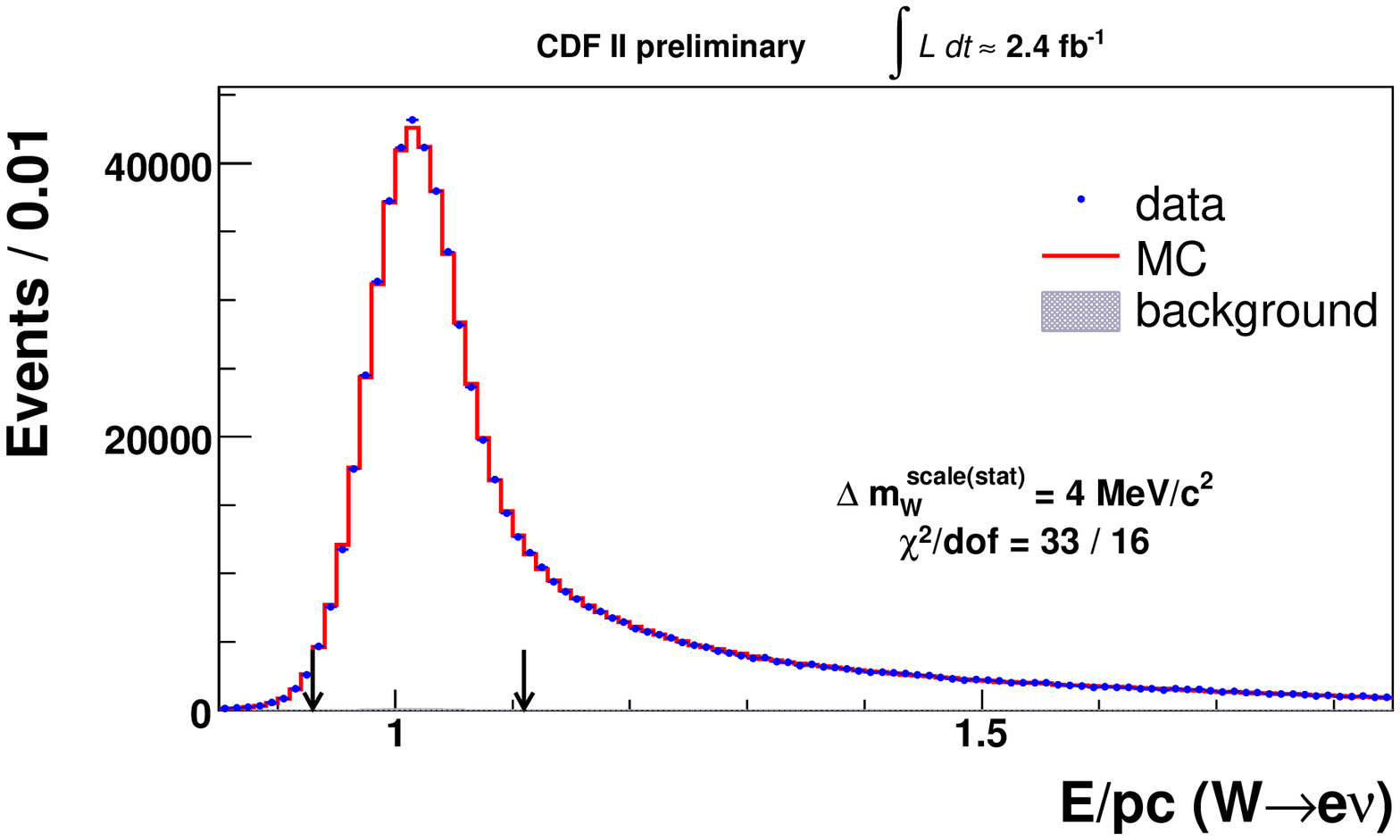}
    \label{eop}
  }
  \caption{
    (a) The invariant mass of the \Zee\ electron pair, calculated
    from the momentum information measured in the COT. (b) The \ep\ 
    distribution of electrons from \Wenu\ decays. The arrows indicate
    the limits of the fitting window.}
\end{figure*}

\begin{figure*}[t]
  \centering
  \subfigure[]{
    \includegraphics[width=80mm]{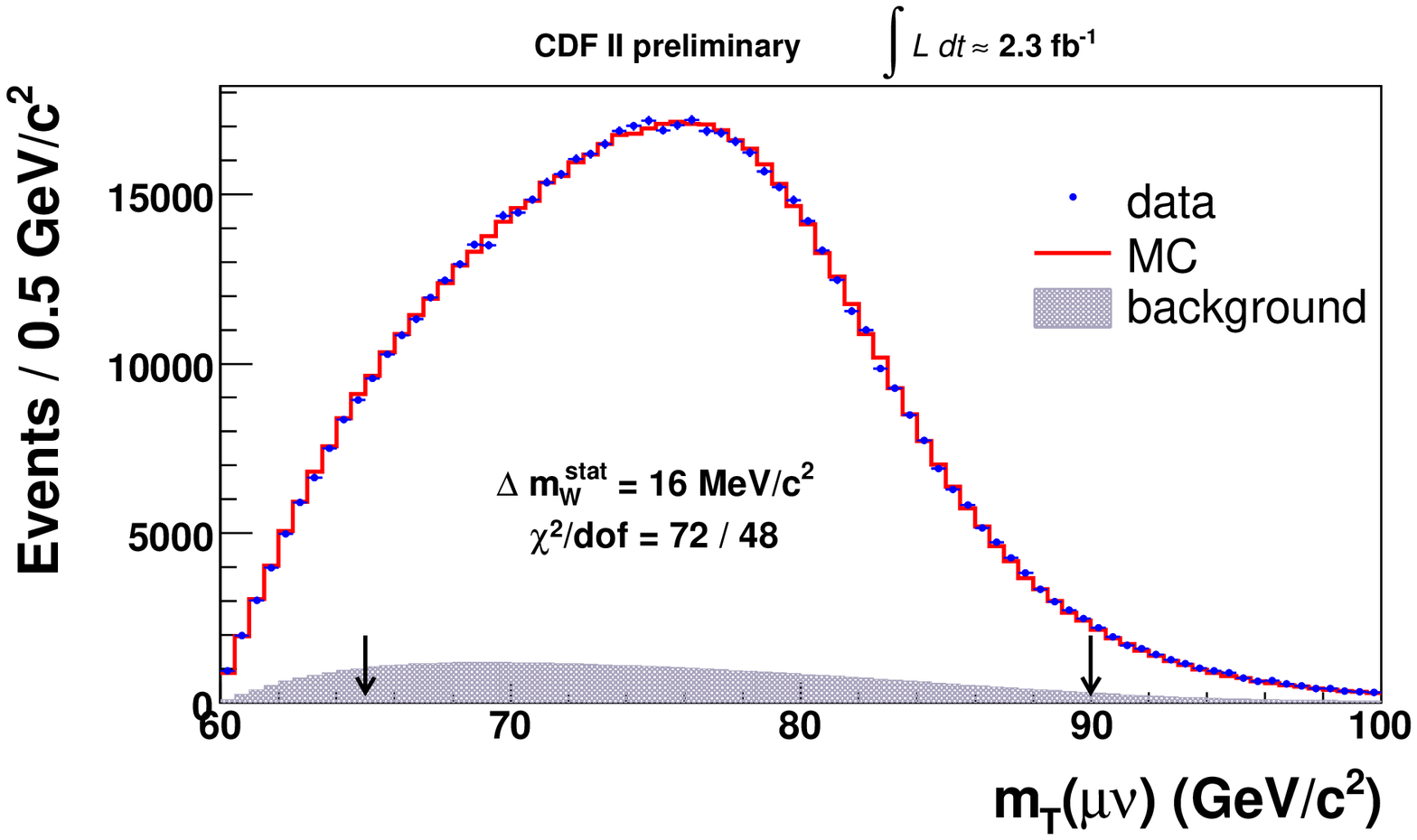}
  }
  \subfigure[]{
    \includegraphics[width=80mm]{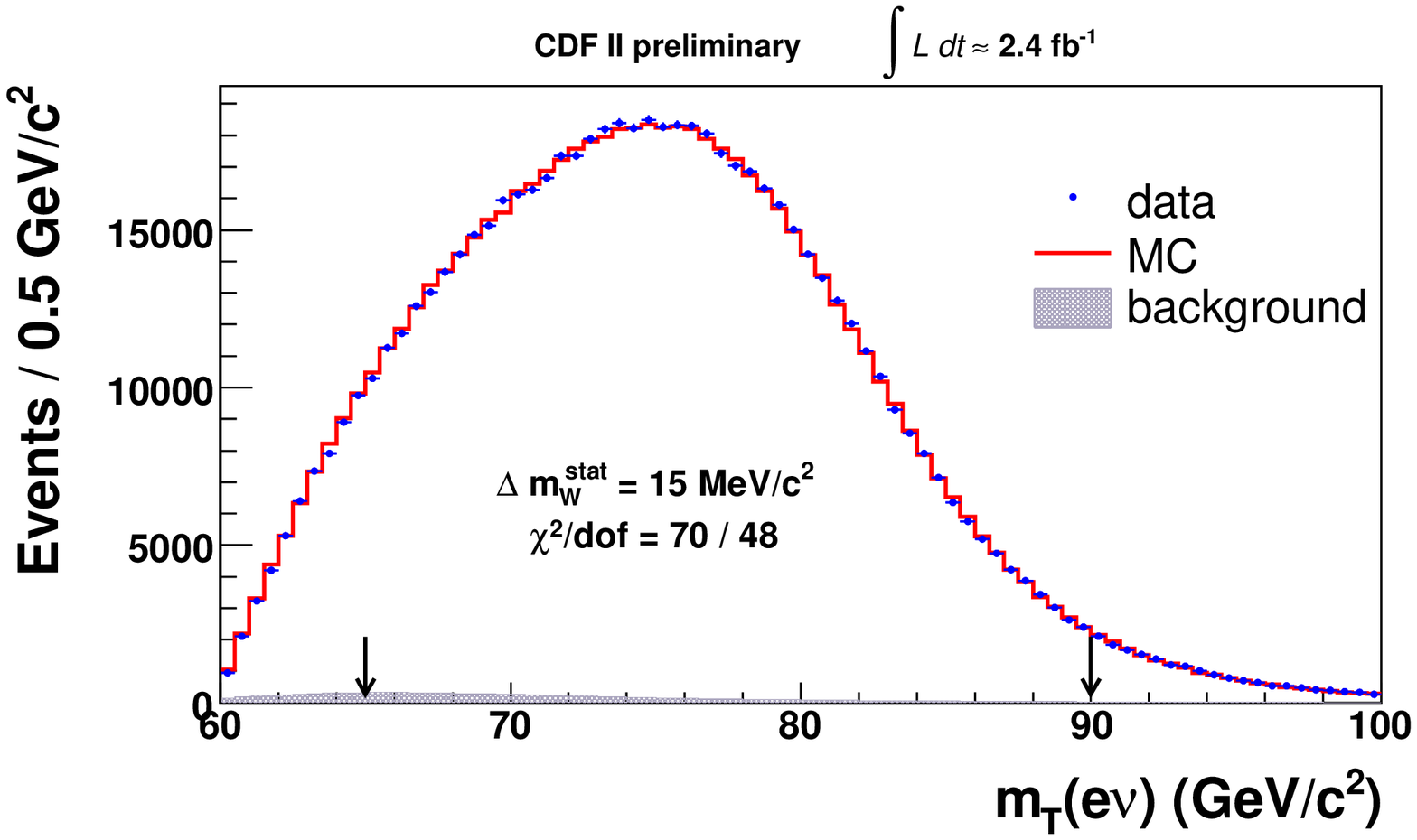}
  }

  \caption{
    The transverse mass distribution for \Wmunu\ (a) and \Wenu\ decays
    (b). The arrows indicate the limits of the fitting window.}
  \label{wmtfit}
\end{figure*}

The statistical uncertainty of the relevant quantity is obtained from
the fits. The statistical uncertainty of the \mw\ fits of \mtr\ and the
\mz\ fits of the invariant dilepton mass is obtained from a fit with
the simulation blinded by an unknown additive offset. In
Table~\ref{statscale} the obtained uncertainties are compared to the
expected statistical uncertainty obtained by scaling the published
result~\cite{wmass} by the square root of the ratio of integrated
luminosities.  The fit uncertainties show that the resolutions of the
distributions in the $2.4~\ifb$ dataset are comparable to those of the
data used in the published result.

\begin{table}[ht]
\begin{center}
\begin{tabular}{|l|c|c|c|}
\hline 
uncertainty     & published & expected     & prelim. fit \\
(all in \mevcd) & 200 \ipb  & 2.3/2.4 \ifb & 2.3/2.4 \ifb\\
\hline 
\dmzstat (\Zmumu)                             & ~43 & 13 & 12 \\
\dmzstat (\Zee)                               & ~67 & 20 & 20 \\
\dmzstat ($\Zee_{\rm track}$)                 & 143 & 42 & 42 \\
\dmwstat (\Wmunu)                             & ~54 & 16 & 16 \\
\dmwstat (\Wenu)                              & ~48 & 14 & 15 \\
\dmwscstat (\ep, \Wenu)                       & ~20 & ~6 & ~5 \\
\dmwscstat (mom. scale, \Jpsimumu -- one bin) & ~20 & ~6 & ~6 \\
\dmwscstat (mom. scale, \Upsmumu)             & ~~5 & ~1 & ~1 \\
\hline 
\end{tabular}
\end{center}
\caption{  
  Statistical scaling, all results in \mevcd. The statistical
  uncertainty for the published result ($200~\ipb$), the expected
  statistical uncertainty obtained by scaling the published result by
  the square root of the ratio of integrated luminosities, and the
  statistical uncertainty obtained from the presented preliminary fit.
}
\label{statscale}
\end{table}


\section{CONCLUSION}

We have published the most precise single measurement of the $W$ boson
mass~\cite{wmass}. The total uncertainty of the published CDF
measurement is smaller than it was projected from previous
measurements, and the improvements in the detector model and the
production and decay model will further reduce the systematic
uncertainties of future measurements. The CDF has set a goal of
measuring the $W$ boson mass with a precision better than that of the
current world average ($25~\mevcd$), using a dataset of approximately
$2.4~\ifb$.

The preliminary plots using $2.4~\ifb$ show that the expected
statistical uncertainty of $W$ mass fits scales with statistics,
indicating that the potential degradation of the level of description
due to the larger spread of instantaneous luminosities and
time-dependent effects has not significantly affected the sensitivity
of the fits.
